\documentclass{PoS}

\title{Parity doubling of nucleons and Delta baryons across the deconfinement phase transition}

\ShortTitle{Parity doubling of $N$ and $\Delta$ baryons across the deconfinement phase transition}

\author{Gert Aarts, Chris Allton, \speaker{Davide De Boni}\thanks{E-mail: d.de-boni.840671@swansea.ac.uk}, Simon Hands, Chrisanthi Praki\\
        Department of Physics, College of Science, Swansea University, Swansea SA2 8PP, United Kingdom}

\author{ Benjamin J\"ager\\
         ETH Z\"urich, Institute for Theoretical Physics, Wolfgang-Pauli-Str. 27, 8093 Z\"urich, Switzerland\thanks{current address}\\
         Department of Physics, College of Science, Swansea University, Swansea SA2 8PP, United Kingdom}

\author{Jon-Ivar Skullerud\\
        Department of Mathematical Physics, National University of Ireland Maynooth, Maynooth, Country Kildare, Ireland}

\abstract{At zero temperature the negative-parity ground states of the nucleon and delta baryons are non-degenerate
with the positive-parity partners due to spontaneous breaking of chiral symmetry. However, chiral symmetry is
expected to be restored at sufficiently high temperature, in particular when going from
the hadronic to the quark-gluon plasma (QGP) phase. This would imply that channels with opposite parity
become degenerate. We study the nucleon (spin $1/2$) and $\Delta$ (spin $3/2$) baryons in both
parity sectors using lattice QCD. The range of temperatures spans both the hadronic and QGP phases. Using the
FASTSUM anisotropic $N_f = 2 + 1$ ensembles, we analyze the correlation functions and the
spectral functions using respectively exponential fits and the Maximum Entropy Method.
We find clear evidence of in-medium effects in the hadronic phase, especially for the negative-parity ground state, and of parity doubling in the QGP phase.}

\FullConference{34th annual International Symposium on Lattice Field Theory\\
		24-30 July 2016\\
		University of Southampton, UK}

\usepackage{color}
\usepackage{bbold}
\usepackage{capt-of}
\usepackage{pstricks}
\usepackage{bm} 
\usepackage{slashed}
\usepackage{booktabs}
\usepackage{tikz}
\usepackage{pgfplots}
\pgfplotsset{/pgf/number format/use comma,compat=newest}

\long\def\comment#1{ }

\newcommand{\beq}{\begin{eqnarray}}
\newcommand{\eeq}{\end{eqnarray}}

\def\x{{\boldsymbol x}}

\newcommand{\avg}[1]{\langle #1 \rangle}

\newcommand{\be}{\begin{eqnarray}}
\newcommand{\ee}{\end{eqnarray}}

\newcommand*\diff{\mathop{}\!\mathrm{d}}

\providecommand*{\eu}{\ensuremath{\mathrm{e}}}

\newcommand{\id}{\mathbb{1}}

\begin{document}

\section{Introduction}

In Nature, at zero temperature the negative-parity ground states of the nucleon and $\Delta$-baryon differ from the positive-parity ones
by about $500$ MeV, a quantity far too big to be explained by the small explicit breaking of chiral symmetry due to the light $u$ and $d$ quarks.
In fact we know that the big difference in mass is due to the spontaneous breaking of chiral symmetry, which is expected to be restored
above the deconfinement temperature, in the quark-gluon plasma phase. One then would expect that a consequence of the restoration
of chiral symmetry is parity doubling.
While there are studies of chiral symmetry at finite temperature in the mesonic sector ~\cite{Rapp}, surprisingly only a few quenched analyses
are available in the baryonic sector ~\cite{Kogut,Datta,Pushkina}. 
This work aims to study parity doubling in the unquenched baryonic sector, in particular for the nucleon and the $\Delta$ particle.
We analyse both correlators and spectral functions below and above the crossover temperature $T_c\,$.
An analysis of correlators in the nucleon sector can be found in \cite{Aarts:2015mma,Aarts:2015xua}.


\section{Baryonic correlators and spectral functions}

The starting point of this work is the baryonic correlator function, which is given in general by (see for instance ~\cite{Montvay:1994cy,Gattringer:2010zz})
\be\label{correlator}
C(x)=\avg{O^\alpha(x)\,\overline{O}^\alpha(0)}\:,
\ee
where the sum over the Dirac index $\alpha$ is implicit. The simplest annihilation operators for the nucleon and $\Delta$-baryon are respectively
\be
O_N^{\alpha}(x) &=&\epsilon_{abc}\,u_a^{\alpha}(x)\left(d^{^T}_b(x)\,C\,\gamma_5\,
u_c(x)\right)\:,
\ee
and ~\cite{Leinweber:2004it,Edwards:2004sx}
\be
O_\Delta^{\alpha}(x) &=&
\epsilon_{abc} \left[ 
2\, u_a^{\alpha}(x)\left(d^{^T}_b(x)\,C\,\gamma_i\, u_c(x)\right) +
d_a^{\alpha}(x)\left(u^{^T}_b(x)\,C\,\gamma_i\, u_c(x)\right) \right]\:,
\ee
where $C$ is the charge conjugation matrix and the spin index $i$ is not summed.
In order to project to a definite parity state we consider the interpolators $O_{N^{\pm}}=P_{\pm}\,O_N$ and $O_{\Delta^{\pm}}=P_{\pm}\,O_\Delta$ in (\ref{correlator}), where
$
P_{\pm} = \frac{1}{2} \left( \id \pm \gamma_4 \right)
$
projects to positive or negative parity.
We consider zero three-momentum correlators $C_{\pm}(\tau)$ by integration over the spatial coordinate $\x\,$.
It is easy to prove that~\cite{Gattringer:2010zz} in the presence of chiral symmetry, a chiral rotation on the quark fields shows
that the two parity channels are degenerate, i.e. $\!C_{\pm}(\tau)=-C_{\mp}(\tau)\,$.
We note that each correlator contains both parity channels, with the positive (negative) one propagating forwards (backwards) in time, i.e. $C_{-}(\tau)=-C_{+}(1/T-\tau)\,$.
Our study comprises also the reconstruction of the spectral functions $\rho(\omega)\,$, obtained using the Maximum
Entropy Method (MEM)~\cite{Asakawa:2000tr}, starting from the spectral relation \cite{wip}
\be
C_{\pm}(\tau)=
\int_{-\infty}^{+\infty}\frac{\diff{\omega}}{2\pi}\,\rho_{\pm}(\omega)\,\frac{\eu^{-\omega\tau}}{1+\eu^{-\omega/T}}\:.
\ee
It can be shown that $\rho_{+}(-\omega)=-\rho_{-}(\omega)\,$, hence negative (positive) frequencies correspond to the negative (positive) parity channel.
In case of parity doubling, $\rho_{\pm}(\omega)$ are even functions.
One can show that $\rho_+(\omega)\geq 0 \quad \forall \omega\,$ ~\cite{wip}, which is an essential requirement for MEM to work.


\section{Lattice setup}

In our study we use the configurations generated by the FASTSUM
collaboration~\cite{Aarts:2014cda,Aarts:2014nba,Amato:2013naa}, with 2+1
flavours of non-perturbatively-improved Wilson fermions. 
The simulation parameters are based on the setup of the Hadron Spectrum Collaboration~\cite{Edwards:2008ja} 
and are summarized in Tab.~\ref{tab:lat}. The configurations and the
correlation functions have been generated using the CHROMA software
package~\cite{Edwards:2004sx} using the SSE
optimizations when possible \cite{McClendon}. The light quark
masses used here give a pion mass of $384(4)\,\mbox{MeV}$~\cite{Lin}, 
whereas the strange quark has been tuned to its physical value.
The anisotropic factor of $a_s/a_\tau = 3.5\,$, with $a_s=0.1227(8)\,\mbox{fm}\,$,
allows us to analyse different temperatures maintaining a large number
of points in the Euclidean time direction. This is in
particular important for reconstructing the spectral function from the hadronic
correlator. The crossover temperature $T_c=183$ MeV has been obtained via the renormalized Polyakov loop.
We employ Gaussian smearing~\cite{Gusken:1989ad} to increase the overlap
with the ground state. To maintain a positive spectral weight, we apply the
smearing on both source and sink, i.e. 
\be
\psi' = \frac{1}{A} \left(\id + \kappa \, H \right)^{n} \psi, 
\ee
where $H$ is the spatial hopping part of the Dirac operator and $A$ an appropriate
normalization. We use a smearing setup of $\kappa = 4.2$ and $n=60\,$, which
was tuned to maximize the length of the plateau for the effective mass of the
ground state for the $24^3 \times 128$ ensemble. The links in the hopping term
are APE smeared~\cite{Albanese:1987ds} using one iteration with $\alpha =
1.33\,$. This smearing procedure is solely applied in the spatial directions and
equally applied on all ensembles and temperatures.

\begin{table}[htbp]
\begin{center}
\begin{tabular}{cccccccc}
\hline
$N_s$ & $N_\tau$  & $T$\,[MeV] & $T/T_c$  & $N_{\rm cfg}$ & $N_{\rm src}$\\
\hline
  24 & 128 & 44	& 0.24 & 139.5 &  16\\
  24 & 40 & 141 & 0.76 & 501   &  4\\
  24 & 36 & 156 & 0.84 & 501   &  4\\
  24 & 32 & 176 & 0.95 & 1000  &  2\\ 
  24 & 28 & 201 & 1.09 & 1001  &  2\\ 
  24 & 24 & 235 & 1.27 & 1001  &  2\\  
  24 & 20 & 281 & 1.52 & 1000  &  2\\ 
  24 & 16 & 352 & 1.90 & 1001  &  2\\ 
\hline
\end{tabular}
\caption{Simulation parameters used in this work. The available statistics
for each ensemble is $N_{\rm cfg} \times N_{\rm src}$. The value $N_{\rm cfg}=139.5$ means that there are $139$ configurations with $16$ sources and $1$ with $8\,$.
The sources were chosen randomly in the four-dimensional lattice.}
\label{tab:lat}
\end{center}
\end{table}


\section{Results for Nucleon and Delta-baryon}

\begin{figure}[t!]
\begin{center}
\includegraphics[width=7.5cm,height=5cm]{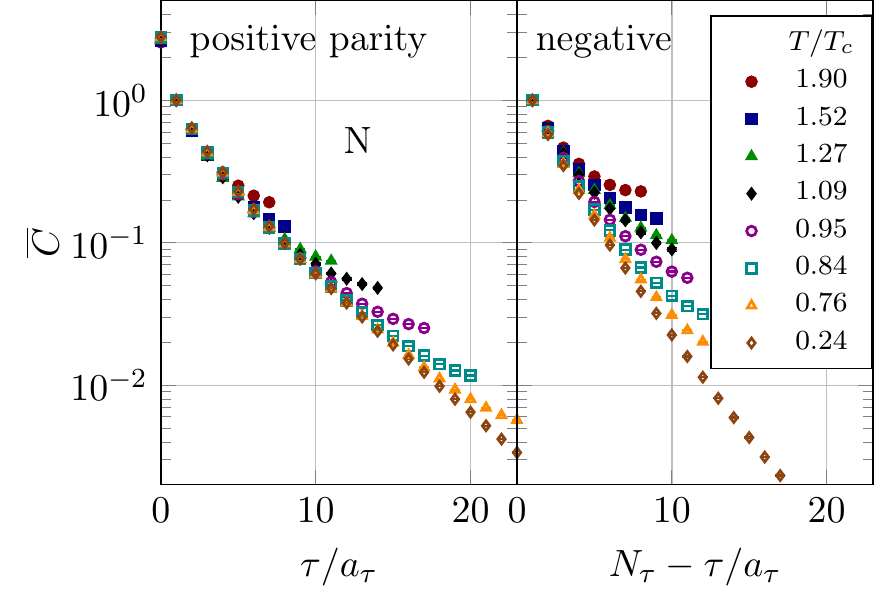}
\includegraphics[width=7.5cm,height=5cm]{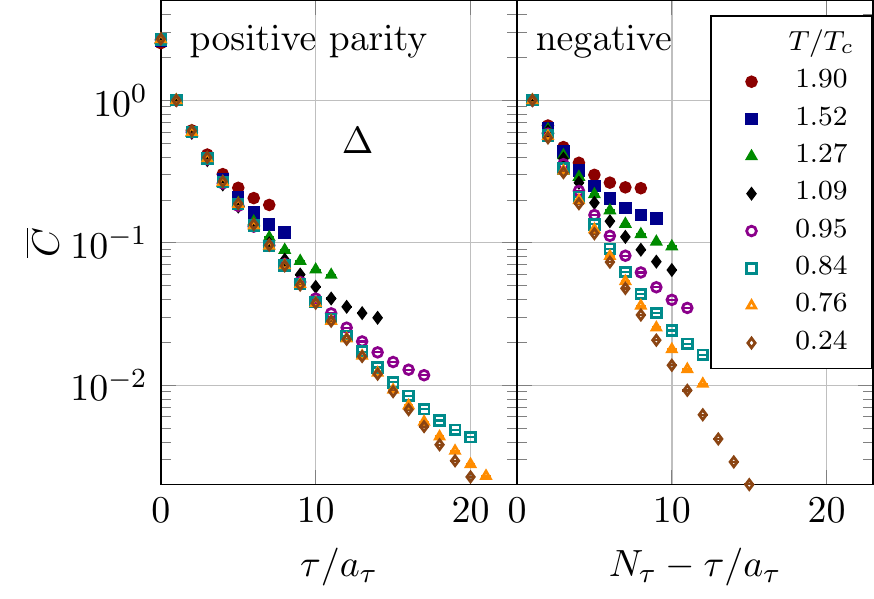}
\caption{On the left (right): Nucleon ($\Delta$-baryon) correlators for the positive and negative parity channels at different temperatures.}
\label{fig:corr}
\end{center}
\end{figure}

\begin{figure}[b!]
\begin{center}
\includegraphics[width=7.5cm,height=5cm]{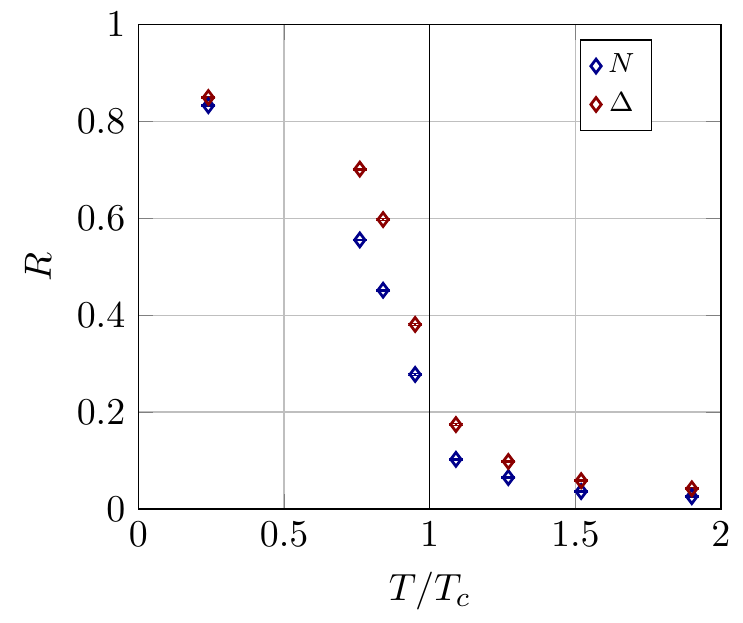}
\includegraphics[width=7.5cm,height=5cm]{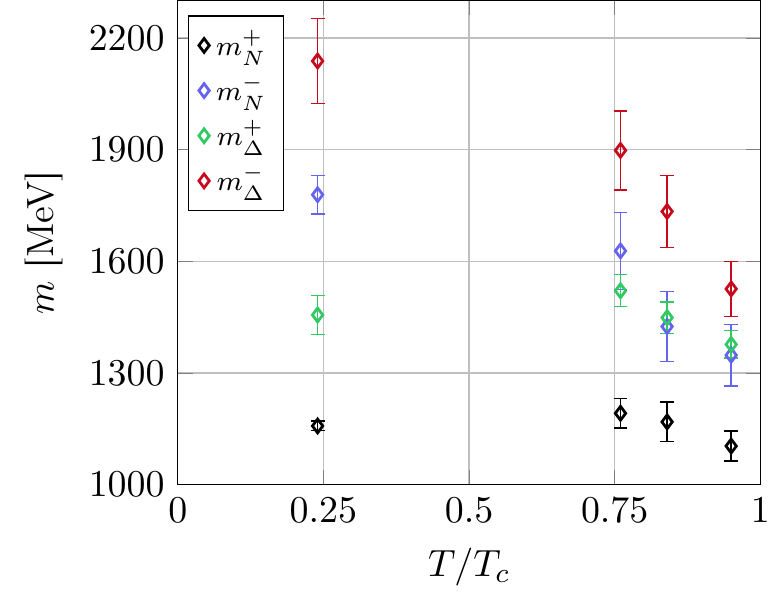}
\caption{On the left: Temperature dependence of the nucleon and $\Delta$-baryon $R$ factors. On the right: Ground state masses obtained using exponential fits to the nucleon and $\Delta$-baryon 
correlators for temperatures below $T_c\,$.}
\label{fig:R&m}
\end{center}
\end{figure}

The correlation functions of the nucleon (spin-$1/2$ baryon octet) and $\Delta$ particle (spin-$3/2$ baryon decuplet) are shown in Fig.~\ref{fig:corr} for both parity partners.  
In order to better compare all data at different temperatures we
normalized the correlation function to the first Euclidean time $\tau=a_\tau$
($\tau=N_\tau a_\tau -a_\tau$) for the positive (negative) parity partner, i.e.  (we write $C=C_+$ for ease of notation)
\be
 \overline{C}_{+}(\tau) = \frac{C\left( \tau \right)}{\langle C(a_\tau) \rangle} \quad
  \mbox{and} \quad
  \overline{C}_{-}(\tau) = \frac{C\left( N_\tau a_\tau- \tau \right)}{\langle C(N_\tau a_\tau - a_\tau)
  \rangle}.
\ee
The negative parity partners exhibit a steeper decay, indicating a heavier mass. Moreover,
both negative parity channels appear more sensitive to temperature.
The left panel of Fig.~\ref{fig:R&m}
shows the summed ratios of the nucleon and $\Delta$-baryon correlators, defined as
\be
\label{eq:ratioR}
R \equiv
\frac{\sum_{n=1}^{N_\tau/
2-1}R(\tau_n)/\sigma^2(\tau_n)}{\sum_{n=1}^{N_\tau/2-1}1/\sigma^2(\tau_n)}\:,\:\:\mbox{with}\quad R(\tau)\equiv \frac{C(\tau)-C(N_\tau a_\tau-\tau)}{C(\tau)+C(N_\tau a_\tau-\tau)}\:.
\ee
By construction the ratio $R$ is between $0$ and $1$. For a completely
symmetric correlation function the ratio vanishes. We use the statistical 
uncertainties as weights in the ratio $R(\tau)$. The plot on the left of
Fig.~\ref{fig:R&m} indicates that the $N$ and $\Delta$ correlation functions become almost symmetric past
the deconfinement transition. This symmetry entails that both parity channels of
the nucleon and $\Delta$-baryon are approximately degenerate above $T_c\,$.
\begin{table}[t!]
\centering
\begin{tabular}{ccccccc} 
$T/T_c$ & $m_N^+$ [MeV] & $m_N^-$ [MeV] & $m_\Delta^+$ [MeV] &
$m_\Delta^-$ [MeV] & $\delta_N$ & $\delta_\Delta$\\ 
\midrule
0.24  & 1158(13)    & 1779(52)   & 1456(53)  & 2138(114) & 0.212(15)     & 0.190(31) \\
0.76  & 1192(39)    & 1628(104)  & 1521(43)  & 1898(106) & 0.155(35)     & 0.110(31) \\
0.84  & 1169(53)    & 1425(94)   & 1449(42)  & 1734(97)  & 0.099(40)     & 0.089(31) \\
0.95  & 1104(40)    & 1348(83)   & 1377(37)  & 1526(74)  & 0.100(35)     & 0.051(28)\\
\end{tabular}
\caption{Ground state masses obtained using exponential fits to the nucleon and $\Delta$-baryon 
correlators for temperatures below $T_c\,$. The masses of the positive and negative parity ground states
include an estimate for statistical and systematic uncertainties.The ratios $\delta_N$ and $\delta_\Delta$ are defined as $\delta=(m_--m_+)/(m_-+m_+)\,$. In
Nature at $T=0$ $\delta_N=0.241$ and $\delta_\Delta=0.160\,$.}
\label{tab}
\end{table}

\begin{figure}[b!]
\begin{center}
\includegraphics[width=7cm,height=5cm]{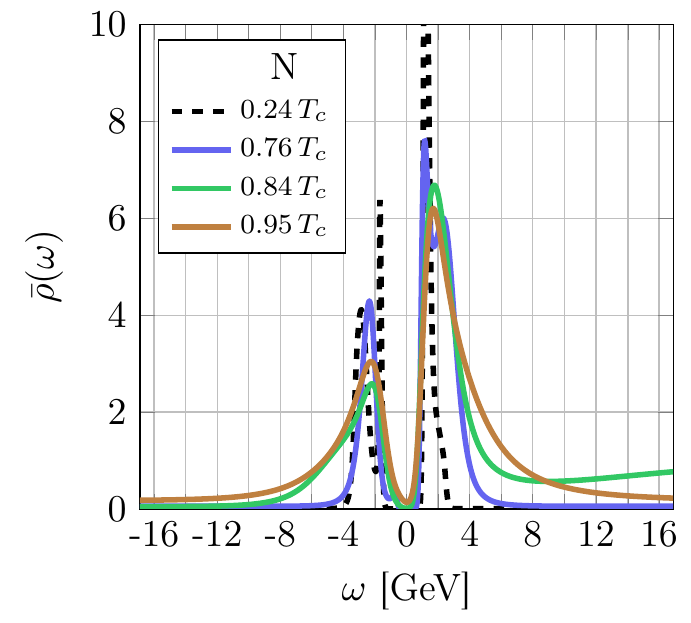}
\includegraphics[width=7cm,height=5cm]{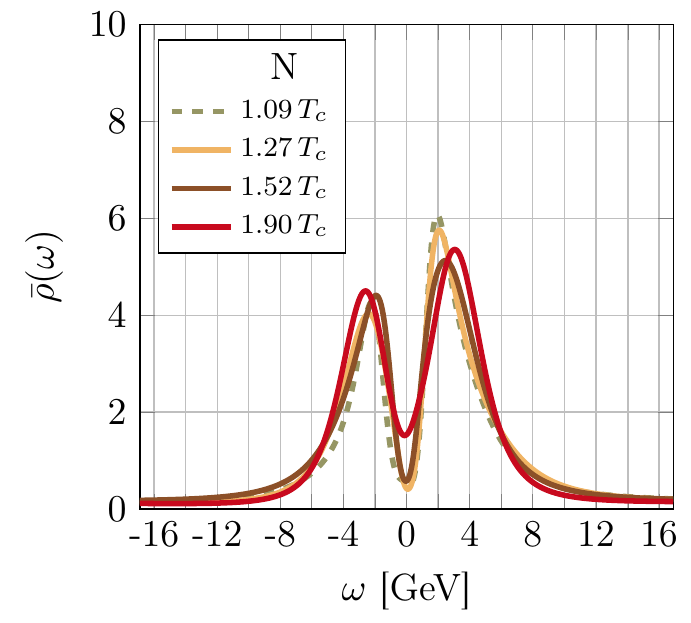}
\includegraphics[width=7cm,height=5cm]{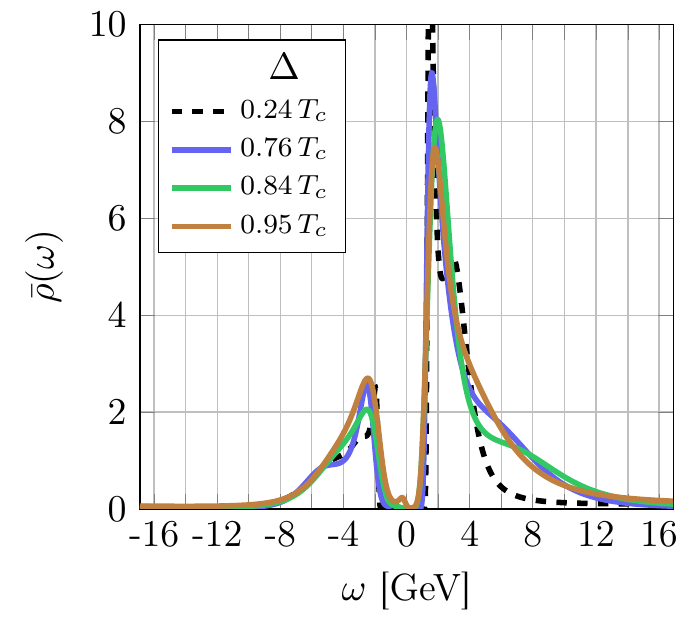}
\includegraphics[width=7cm,height=5cm]{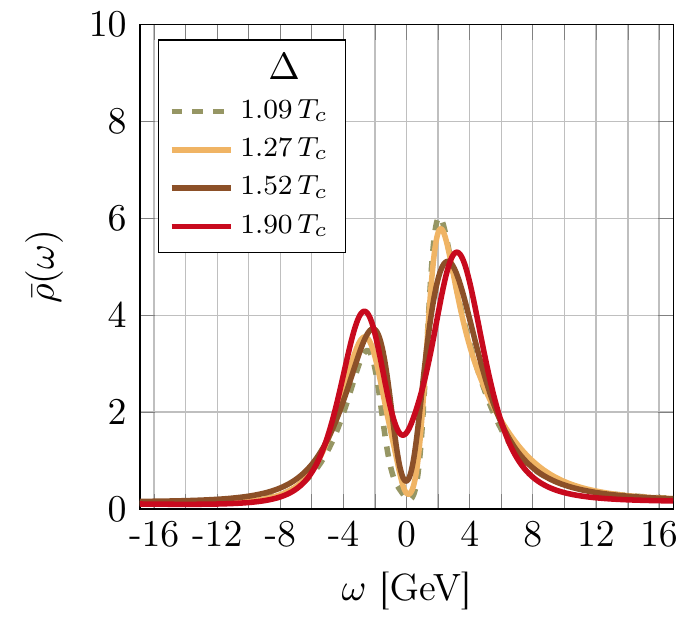}
\caption{On the left (right): Normalised spectral functions obtained using MEM below (above) $T_c$ for the nucleon (top) and $\Delta$-baryon (bottom), where the dimensionless object $\bar\rho(\omega)$ is obtained from the normalised correlator $C(\tau)/(a_\tau\,\langle C(\tau=0)\rangle)\,$.}
\label{delta-spec}
\end{center}
\end{figure}
To extract the ground state masses in the confined phase (see right panel of Fig.~\ref{fig:R&m} and Tab.~\ref{tab}), we fit the correlation function to
a simple exponential for the positive and negative ground state, i.e.
\be
C_+(\tau)= A_{+}\,\eu^{-m_{+}\tau}+A_{-}\,\eu^{-m_{-}(N_\tau a_\tau-\tau)}\:. 
\label{eq:ansatz}
\ee
We vary the Euclidean time interval, excluding very small Euclidean times to
suppress excited states further, to assess the systematic uncertainties of all
four fit parameters. We apply the so-called Extended Frequentist
Method~\cite{Yao:2006px,Durr}, which takes all variation into account and weights
the final results in respect to the obtained $p$-value. More details on this
method can be found in~\cite{Yao:2006px,Durr}.\\
We observe that the ground state mass in the positive-parity channel is approximately independent of the temperature,
whereas the ground state in the negative-parity channel becomes lighter as the temperature is increased.
Therefore in-medium effects are stronger in the negative-parity channel and coincide with
the restoration of chiral symmetry.


The corresponding spectral functions are shown in Fig.~\ref{delta-spec}.
We recall that the spectral function in the positive-parity channel is shown at $\omega>0\,$, and in the negative-parity 
channel at $\omega<0\,$. Below $T_c$ the spectral functions are asymmetric, meaning that
the positive- and negative-parity channels are distinct. However, above $T_c$ we see a clear signal
of parity restoration since the spectral function becomes more symmetric as temperature increases. We do not show the error bars
of the spectral functions in order to clearly visualise their temperature dependence.
However, error bars do not change the above conclusions.

\section{Conclusions}

By studying both the baryonic correlators and spectral functions, this work shows that there is a clear signal of parity doubling across the
deconfinement transition in the ground state for both the nucleon and $\Delta$-baryon.
This can be seen both from the almost vanishing $R$ factor and from the correlators and spectral functions.
Since the asymmetry in the parity partners at zero temperature is mainly due to spontaneous breaking of chiral symmetry, the observed parity doubling
can be understood from the restoration of chiral symmetry, which is expected to occur at high temperature.
Notice that there is still a small explicit breaking of chiral symmetry since we are using
massive $u$ and $d$ quarks in the Wilson formulation.


\section{Acknowledgements}

We acknowledge
the Science Foundation Ireland, grant 11-RFP.1-PHY-3201 (SMR),
the grant NRF- 2015R1A2A2A01005916 (SK),
the STFC grants ST/L000369/1,            
ST/K000411/1, ST/H008845/1, ST/K005804/1 and ST/K005790/1,     
the STFC DiRAC HPC Facility (www.dirac.ac.uk),
PRACE grants 2011040469 and
Pra05 1129q,
CINECA grant INF14 FTeCP (CINECA-INFN agreement).\\
We are grateful to the Royal Society, the Wolfson Foundation and the Leverhulme Trust for support.



\newpage

\begin{thebibliography}{99}

\bibitem{Rapp}
R.~Rapp and J.~Wambach, {\emph{Adv. Nucl. Phys.} \bf{25 (2000) 1}},
[\href{http://arxiv.org/abs/hep-ph/9909229}{{\tt
  hep-ph/9909229}}].

\bibitem{Kogut}
C.~E.~DeTar and J.~B.~Kogut, {\emph{Phys. Rev. Lett.} \bf{59 (1987) 399}}; {\emph{Phys. Rev. D} \bf{36 (1987) 2828}}.

\bibitem{Datta}
S. Datta, S. Gupta, M. Padmanath, J. Maiti and N. Mathur,
{\emph{JHEP} \bf{1302 (2013) 145}}
[\href{http://arxiv.org/abs/1212.2927}{{\tt
  1212.2927}}].

\bibitem{Pushkina}
I. Pushkina et al. [QCD-TARO Collaboration], {\emph{Phys. Lett. B} \bf{609 (2005) 265}},
[\href{http://arxiv.org/abs/hep-lat/0410017}{{\tt
  hep-lat/0410017}}].

\bibitem{Aarts:2015mma}
G.~Aarts, C.~Allton, S.~Hands, B.~J\"ager, C.~Praki and J.-I. Skullerud,
  \href{http://dx.doi.org/10.1103/PhysRevD.92.014503}{\emph{Phys. Rev.} {\bf
  D92} (2015) 014503}, [\href{http://arxiv.org/abs/1502.03603}{{\tt
  1502.03603}}].

\bibitem{Aarts:2015xua}
G.~Aarts, C.~Allton, S.~Hands, B.~J\"ager, C.~Praki and J.-I. Skullerud,
 {\emph{PoS} {\bf LATTICE2015} (2015) 183}, [\href{http://arxiv.org/abs/1510.04040}{{\tt
  1510.04040}}].

\bibitem{Montvay:1994cy}
I.~Montvay and G.~Munster,
\newblock Cambridge University Press, 1997.

\bibitem{Gattringer:2010zz}
C.~Gattringer and C.~B. Lang,
  \href{http://dx.doi.org/10.1007/978-3-642-01850-3}{\emph{Lect. Notes Phys.}
  {\bf 788} (2010) 1--343}.

\bibitem{Leinweber:2004it}
D.~B. Leinweber, W.~Melnitchouk, D.~G. Richards, A.~G. Williams and J.~M.
  Zanotti,
  \href{http://dx.doi.org/10.1007/11356462_4}{\emph{Lect. Notes Phys.} {\bf
  663} (2005) 71--112}, [\href{http://arxiv.org/abs/nucl-th/0406032}{{\tt
  nucl-th/0406032}}].

\bibitem{Edwards:2004sx}
{\scshape SciDAC, LHPC, UKQCD} collaboration, R.~G. Edwards and B.~Joo,
  \href{http://dx.doi.org/10.1016/j.nuclphysbps.2004.11.254}{\emph{Nucl. Phys.
  Proc. Suppl.} {\bf 140} (2005) 832},
  [\href{http://arxiv.org/abs/hep-lat/0409003}{{\tt hep-lat/0409003}}].


\bibitem{Asakawa:2000tr}
M.~Asakawa, T.~Hatsuda and Y.~Nakahara,
  \href{http://dx.doi.org/10.1016/S0146-6410(01)00150-8}{\emph{Prog. Part.
  Nucl. Phys.} {\bf 46} (2001) 459--508},
  [\href{http://arxiv.org/abs/hep-lat/0011040}{{\tt hep-lat/0011040}}].



\bibitem{wip}
G.~Aarts, C.~Allton, D.~De Boni, S.~Hands, B.~J\"ager, C.~Praki and J.-I. Skullerud, \emph{{Work in progress}}.



\bibitem{Aarts:2014cda}
G.~Aarts, C.~Allton, T.~Harris, S.~Kim, M.~P. Lombardo, S.~M. Ryan et~al.,
  \href{http://dx.doi.org/10.1007/JHEP07(2014)097}{\emph{JHEP}
  {\bf 07} (2014) 097}, [\href{http://arxiv.org/abs/1402.6210}{{\tt
  1402.6210}}].

\bibitem{Aarts:2014nba}
G.~Aarts, C.~Allton, A.~Amato, P.~Giudice, S.~Hands and J.-I. Skullerud,
  \href{http://dx.doi.org/10.1007/JHEP02(2015)186}{\emph{JHEP} {\bf
  02} (2015) 186}, [\href{http://arxiv.org/abs/1412.6411}{{\tt 1412.6411}}].

\bibitem{Amato:2013naa}
A.~Amato, G.~Aarts, C.~Allton, P.~Giudice, S.~Hands and J.-I. Skullerud,
  \href{http://dx.doi.org/10.1103/PhysRevLett.111.172001}{\emph{Phys. Rev.
  Lett.} {\bf 111} (2013) 172001}, [\href{http://arxiv.org/abs/1307.6763}{{\tt
  1307.6763}}].

\bibitem{Edwards:2008ja}
R.~G. Edwards, B.~Joo and H.-W. Lin,
  \href{http://dx.doi.org/10.1103/PhysRevD.78.054501}{\emph{Phys. Rev.} {\bf
  D78} (2008) 054501}, [\href{http://arxiv.org/abs/0803.3960}{{\tt
  0803.3960}}].

\bibitem{McClendon}
C. McClendon,
\href{http://www.jlab.org/~edwards/qcdapi/reports/dslash_p4.pdf}{\emph{Jlab preprint, JLAB-THY-01-29}}.

\bibitem{Lin}
H.-W. Lin \textit{et al.},
  \href{http://dx.doi.org/10.1103/PhysRevD.79.034502}{\emph{Phys. Rev.} {\bf
  D79} (2009) 034502}, [\href{http://arxiv.org/abs/0810.3588}{{\tt
  0810.3588}}].

\bibitem{Gusken:1989ad}
S.~Gusken, U.~Low, K.~H. Mutter, R.~Sommer, A.~Patel and K.~Schilling,
  \href{http://dx.doi.org/10.1016/S0370-2693(89)80034-6}{\emph{Phys.
  Lett.} {\bf B227} (1989) 266--269}.

\bibitem{Albanese:1987ds}
{\scshape APE} collaboration, M.~Albanese et~al.,
  \href{http://dx.doi.org/10.1016/0370-2693(87)91160-9}{\emph{Phys. Lett.} {\bf
  B192} (1987) 163--169}.


\bibitem{Yao:2006px}
{\scshape Particle Data Group} collaboration, W.~M. Yao et~al.,
  \href{http://dx.doi.org/10.1088/0954-3899/33/1/001}{\emph{J. Phys.} {\bf G33}
  (2006) 1--1232}.

\bibitem{Durr}
S. Durr et al.,
\href{http://arxiv.org/abs/0906.3599}{\emph{Journal Science} {\bf 322} (2008) 1224-1227}
[\href{http://arxiv.org/abs/0906.3599}{{\tt 0906.3599}}]






\end{thebibliography}
\end{document}